\documentclass[
 aip,
 amsmath,amssymb,
preprint,
 reprint,
]{revtex4-1}

\usepackage{graphicx}% Include figure files
\usepackage{dcolumn}% Align table columns on decimal point
\usepackage{bm}% bold math
\usepackage[utf8]{inputenc}
\usepackage[T1]{fontenc}
\usepackage{siunitx}
\usepackage{mathptmx}
\usepackage{etoolbox}

\makeatletter
\def\@email#1#2{%
 \endgroup
 \patchcmd{\titleblock@produce}
  {\frontmatter@RRAPformat}
  {\frontmatter@RRAPformat{\produce@RRAP{*#1\href{mailto:#2}{#2}}}\frontmatter@RRAPformat}
  {}{}
}%
\makeatother

\begin{document}

\preprint{AIP/123-QED}

\title{Electronic transport mechanisms in a thin crystal of the Kitaev candidate $\alpha$-RuCl$_3$ probed through guarded high impedance measurements}
%Electronic transport mechansms in a thin crystal of the Kitaev candidate $\alpha$-RuCl$_3$ probed through guarded high impedance measurements
%Efros-Shklovskii variable range hopping in the spin-orbit assisted Mott insulator $\alpha$-RuCl$_3$

\author{Patrick Barfield} 
\affiliation{Department of Physics and Astronomy, California State University Long Beach, Long Beach, California 90840, USA}

\author{Vinh Tran} 
\affiliation{Department of Physics and Astronomy, California State University Long Beach, Long Beach, California 90840, USA}

\author{Vikram Nagarajan}
\affiliation{Department of Physics, University of California Berkeley, Berkeley, CA, 94720}

\author{Maya Martinez} 
\affiliation{Department of Physics and Astronomy, California State University Long Beach, Long Beach, California 90840, USA}

\author{Amirari Diego} 
\affiliation{Department of Physics and Astronomy, California State University Long Beach, Long Beach, California 90840, USA}

\author{Derek Bergner} 
\affiliation{Department of Physics and Astronomy, California State University Long Beach, Long Beach, California 90840, USA}

\author{Alessandra Lanzara}
\affiliation{Department of Physics, University of California Berkeley, Berkeley, CA, 94720}
\affiliation{Materials Sciences Division, Lawrence Berkeley National Laboratory, Berkeley,
California 94720, United States}

\author{James G. Analytis}
\affiliation{Department of Physics, University of California Berkeley, Berkeley, CA, 94720}
\affiliation{Materials Sciences Division, Lawrence Berkeley National Laboratory, Berkeley,
California 94720, United States}

\author{Claudia Ojeda-Aristizabal}
\affiliation{Department of Physics and Astronomy, California State University Long Beach, Long Beach, California 90840, USA}
\email[Corresponding author~]{Claudia.Ojeda-Aristizabal@csulb.edu}

\date{\today}

\begin{abstract}
    $\alpha$-RuCl$_3$ is considered to be the top candidate material for the experimental realization of the celebrated Kitaev model, where ground states are quantum spin liquids (QSL) with interesting fractionalized excitations. It is however known that additional interactions beyond the Kitaev model trigger in $\alpha$-RuCl$_3$, a long-range zigzag antiferromagnetic ground state. 
    %due to its honeycomb lattice and bond directional exchange interactions, is of interest as a possible realization of the Kitaev-Heisenberg model exhibiting quantum spin liquid in ground states. 
    In this work, we investigate a nanoflake of $\alpha$-RuCl$_3$ through guarded high impedance measurements aimed at reaching through electronic transport, the regime where the system turns into a zigzag antiferromagnet. 
    %improving the signal to noise ratio in highly insulating material. 
    We investigated a variety of temperatures (\SI{1.45}{\kelvin} - \SI{175}{\kelvin}) and out-of-plane magnetic fields ranging up to \SI{11}{\tesla}. We found a clear signature of a structural phase transition at $\approx 160$\,K as reported for thin crystals of $\alpha$-RuCl$_3$, as well as a thermally activated behavior at temperatures above $\approx 30$\,K with a characteristic activation energy significantly smaller than the energy gap that we observe for $\alpha$-RuCl$_3$ bulk crystals through our Angle Resolved Photoemission Spectroscopy (ARPES) experiments. Additionally we found that below $\approx 30$\,K,  transport is ruled by Efros-Shklovskii (ES) VRH. These observations point to the presence of Coulomb impurities in our thin crystals. Most importantly, our data shows that below the magnetic ordering transition known for bulk $\alpha$-RuCl$_3$ ($\approx 7$\,K), there is a clear deviation from VRH or thermal activation transport mechanisms. 
    %finding that there is a clear deviation from transport mechanisms such as variable range hopping (VRH) or thermal activation at temperatures below the magnetic ordering transition. In particular, we find that above that temperature ($\approx 7$\,K), transport is ruled by Efros-Shklovskii (ES) VRH, becoming thermally activated at $\approx 30$\,K and both mechanisms being suppressed below the critical temperature. ES-VRH is consistent with the zero density of states at the Fermi level that we have observed through our Angle Resolved Photoemission Spectroscopy (ARPES) measurements. 
    Our work demonstrates the possibility of reaching through specialized high impedance measurements, the thrilling ground states predicted for $\alpha$-RuCl$_3$ at low temperatures in the frame of the Kitaev model, and informs about the transport mechanisms in this material in a wide temperature range as well as on important characteristic quantities such as the localization length of the impurities in a thin $\alpha$-RuCl$_3$ crystal.  
    %(u) transport mechanism at  We find that although there is no dependence of the electronic transport on out-of-plane magnetic fields,  transport is dominated primarily by thermal activation for high temperatures (30-85\,K) and low bias (25\,mV). On the other hand, for low temperatures (10-30\,K) and low bias, $\ln(I)$ exhibits a $T^{-1/2}$ dependence, suggesting the transport is dominated by Efros-Shklovskii variable range hopping. Finally, we find that for high bias and, the transport best fits a temperature and electric field dependent variable range hopping.  
\end{abstract}

\pacs{}

\maketitle

\section{Introduction}
\begin{figure} 
\includegraphics[width=0.45\textwidth]{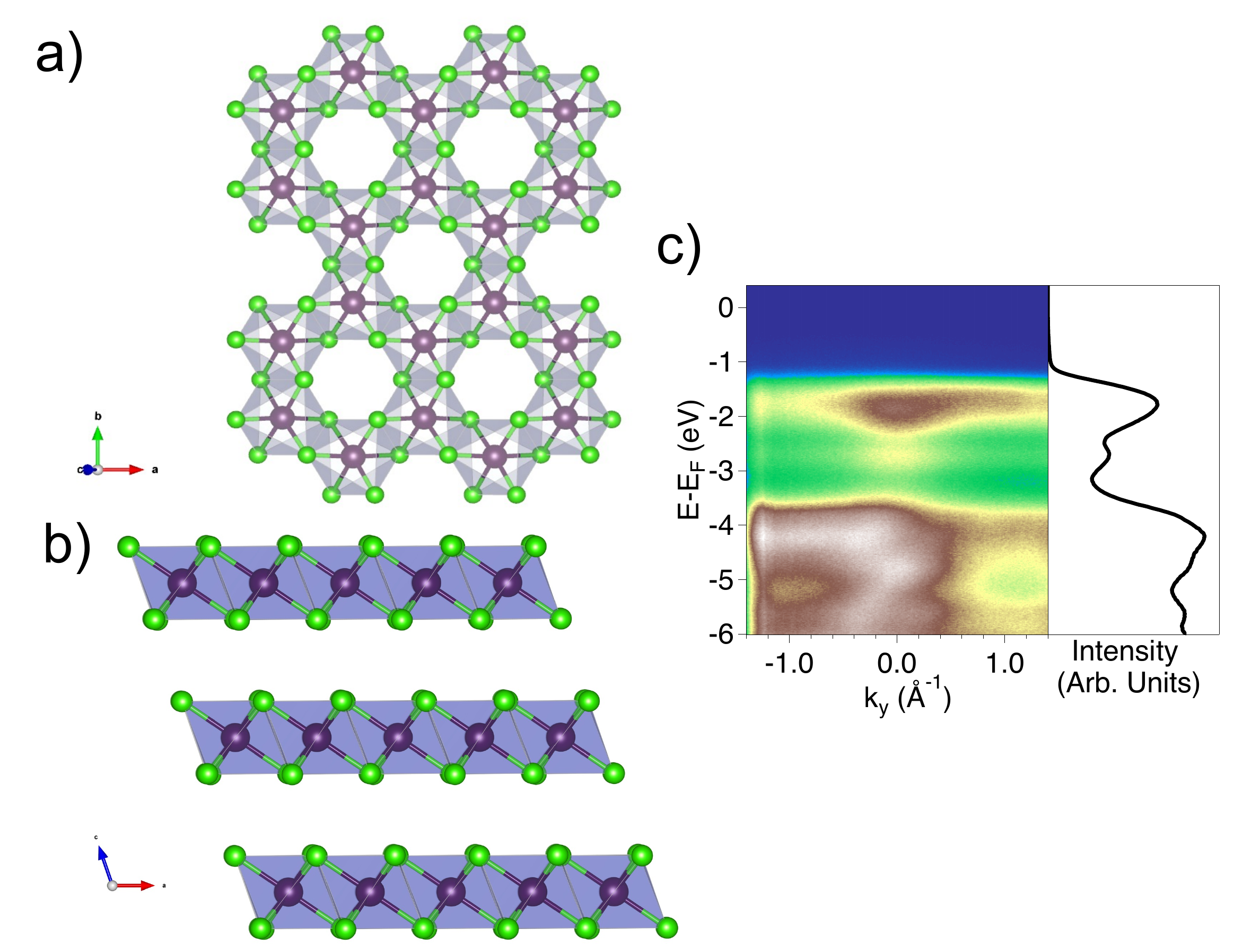}
\caption{\textbf{Crystal and electronic structure of $\alpha$-RuCl$_3$}\\(a) Crystal structure of $\alpha$-RuCl$_3$ where Ru$^{3+}$ ions form a honeycomb layer, with each ion surrounded by a Cl octahedral cage sharing an edge with its neighbors. (b) Low temperature C2/m arrangement of the honeycomb layers in an ABC configuration (see text) 
(c) Measured electronic band structure in a k direction across $\Gamma$ within the first Brillouin zone. Momentum integrated (in a range of -1.0\,\AA$^{-1}$ $\leq k_y \leq$ 1.0\,\AA$^{-1}$ ) energy distribution curve showing that the density of states vanishes between the top of the valence band and the the Fermi energy.} 
\label{fig:figure1}
\end{figure}

$\alpha$-RuCl$_3$ is a van der Waals material predicted to host long-desired Kitaev quantum excitations, of interest for fault-tolerant topological quantum computing \cite{PhysRevB.90.041112, PhysRevB.91.241110, Takagi2019}. In $\alpha$-RuCl$_3$, Ru$^{3+}$ ions form a honeycomb lattice (Fig.\ref{fig:figure1}a), where each ion is surrounded by an octahedral arrangement of Cl atoms. Neighboring octahedra are edge sharing, which according to the theoretical model by Jackeli and Khaliullin leads to dominant Kitaev interactions between spin-orbit entangled $j=1/2$ moments of the Ru$^{3+}$ ions, in contrast to corner sharing octahedra, that result in traditional symmetric Heisenberg interactions \cite{PhysRevLett.102.017205}. While honeycomb layers are bounded by van der Waals interactions, $\alpha$-RuCl$_3$ has at room temperature a 3-dimensional structure with a unit cell formed by three honeycomb layers stacked in an AB configuration forming a rhombohedral crystal structure described by the P3$_1$12 space group \cite{Ziatdinov2016} (see Suppl. Mat.). In the absence of crystal deformations, at low temperatures (below 155\,K), the unit cell becomes monoclinic described by the C2/m space group\cite{PhysRevB.93.134423} (see Fig. \ref{fig:figure1} b). In such a configuration, the system presents a 7\,K transition to a magnetic order, where alternating chains of ferromagnets run along the zigzag direction of the honeycomb, referred to as a zig-zag antiferromagnetic order, with the magnetic moments oriented $\pm35^{\circ}$ from the ab plane \cite{PhysRevB.93.134423}. It has been demonstrated through temperature-dependent specific heat capacity and single-crystal neutron diffraction experiments, that crystal deformations may lead to stacking faults and an additional, broader magnetic ordered transition at 14\,K, attributed to a two-layer AB stacking order \cite{PhysRevB.93.134423}. 

Overall, the presence of a long-range magnetic order state in $\alpha$-RuCl$_3$ derives from the existence of additional interactions beyond the Kitaev model. The true Hamiltonian for $\alpha$-RuCl$_3$ is still controversial, but it is believed to be formed, in addition to a bond-dependent Kitaev ferromagnetic term ($K<0$), of an isotropic Heisenberg term (J) that includes second and third order-neighbor interactions, as well as of off-diagonal antiferromagnetic $\Gamma$ terms ($\Gamma>0$) that are as the Kitaev exchange, bond-directional. It has been demonstrated through resonant  x-ray scattering \cite{Sears2020} that the $\Gamma$ term, that keeps the ordered momentum in the a-c plane \cite{PhysRevB.94.064435}, is comparable to the Kitaev term $K$ and responsible for the observed large anisotropy of the magnetic susceptibility in $\alpha$-RuCl$_3$ \cite{PhysRevB.91.144420,PhysRevB.92.235119,PhysRevB.91.094422,PhysRevB.91.180401}. Outcome of these multiple interactions in $\alpha$-RuCl$_3$, is the previously mentioned zig-zag antiferromagnetic ground state that can be conveniently suppressed by applying a $\approx$7\,T in-plane magnetic field along the zig-zag direction a (see Fig \ref{fig:figure1}a), yielding a quantum spin liquid (QSL) \cite{Kasahara2018, Czajka2021} and bringing experimentalists a step closer to the desired anyonic excitations predicted in the frame of the Kitaev model \cite{osti_20766977}. Much stronger fields are needed in the c direction (up to $\approx$ 33\,T) to destroy the zig-zag antiferromagnetic order and reaching a quantum spin liquid state, as recently reported  \cite{https://doi.org/10.48550/arxiv.2201.04597}.

From an electronic point of view, $\alpha$-RuCl$_3$ is a transition metal oxide with partially filled 4d shells (a halide). Despite the expected large atomic overlap that should lead to large electronic bandwidths and a metallic behavior, important spin-orbit coupling triggers the formation of separate $j=3/2$ and $j=1/2$ bands, leading regardless of largely suppressed electronic correlations, to a Mott insulating state \cite{TREBST20221, PhysRevB.90.041112}.  

Here, we study through electronic transport measurements at low temperatures a $\alpha$-RuCl$_3$ thin crystal device, finding the different electronic transport mechanisms that rule this Kitaev candidate material in a wide range of temperatures. %Our study ratifies the importance of the moderate electronic correlations in this 4d transition metal halide 
Our study puts in evidence the influence of charged impurities in $\alpha$-RuCl$_3$ thin crystal devices and shows a clear change in the electron transport mechanism below the zig-zag magnetic order temperature known for the bulk.

\section{Sample Fabrication and Methods} 
\begin{figure}
    \includegraphics[width=0.45\textwidth]{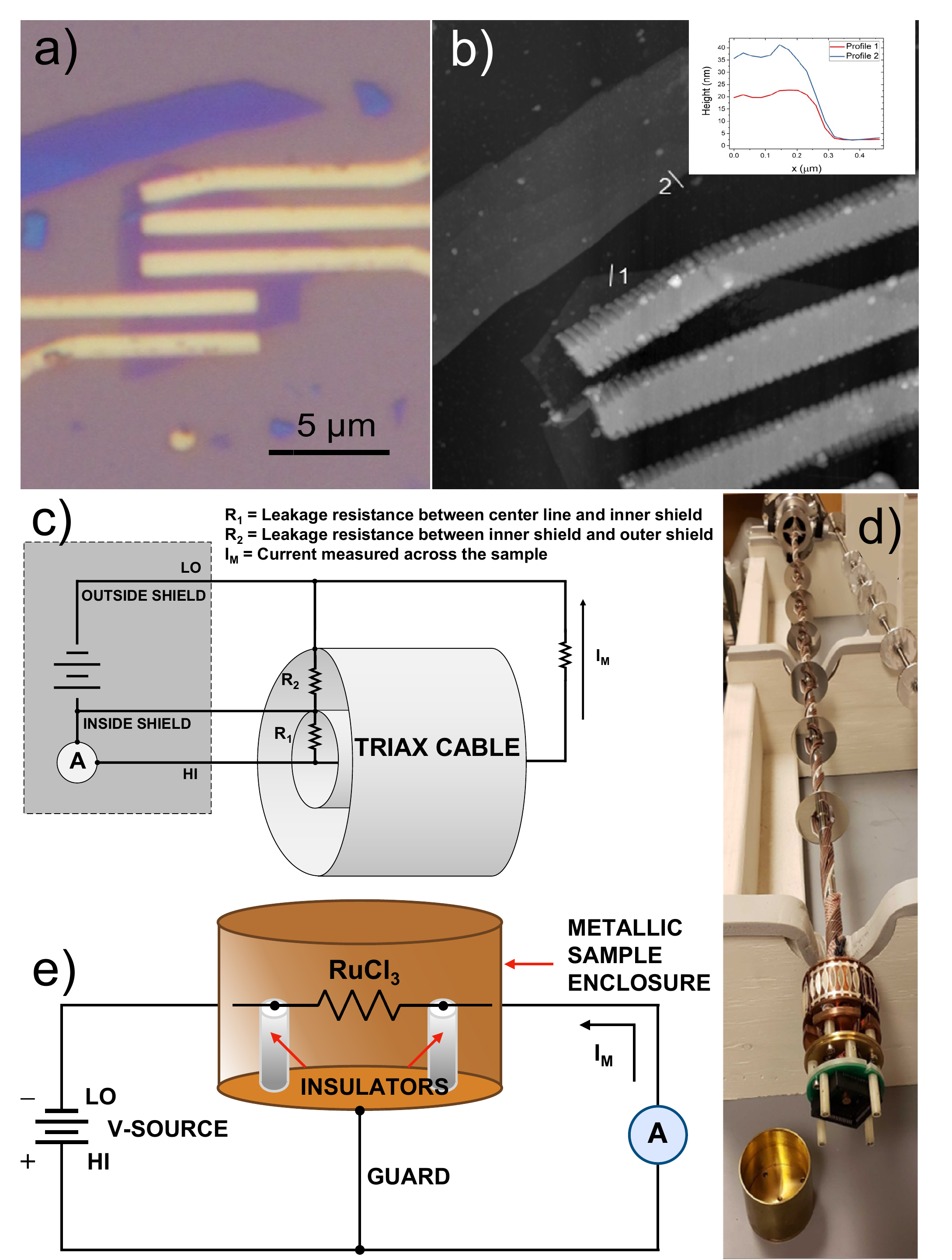}\caption{\textbf{Fabricated device and Experimental Setup}\\ 100x optical (a) and atomic force microscopy (AFM) (b) images of the completed $\alpha$-RuCl$_3$ device. Inset shows measured height of $\alpha$-RuCl$_3$ nanoflake used for the device alongside a nearby, thicker nanoflake for comparison. (c) Circuit schematic of the guarded force-voltage-measure-current setup used in this experiment making use of triax lines, wired to a custom probe with a copper sample enclosure (d). (e) Schematic showing the connection of the guard to the sample enclosure.} 
    %This further isolates the measured current from leakage currents through the enclosure to ground as well as from electromagnetic interference. 
    \label{fig:figure2}
    \end{figure}

%Sample Fabrication\Method%
    High quality crystals of $\alpha$-RuCl$_3$ were grown using the sublimation physical vapor transport method as outlined in \onlinecite{May2020-pa}. About 0.2\,g of commercial RuCl$_3$ powder (Alfa Aesar) was sealed in a quartz ampoule under vacuum and placed in a 2-zone furnace. The source and sink temperatures were 700\,C and 600\,C, respectively; these temperatures were held for 20 days, with single crystals forming at the cold end. The obtained crystals were mechanically exfoliated onto a Si/SiO$_3$ (285\,nm) substrate. Once a suitable thin nanoflake ($\sim$10-20\,nm) was identified via optical and atomic force microscopy, the sample was coated with poly(methyl methacrylate) 495 and 950 electron beam resist. An electrode pattern was drawn using standard electron beam lithography and a thin film of Ti/Au (10 nm/100 nm) was deposited  using an electron beam evaporator. The $\alpha$-RuCl$_3$ nanoflake used in this experiment was verified by AFM to be $\sim$18\,nm thick (Fig. \ref{fig:figure2}(b)) which corresponds to $\sim$15 layers. 
    % \color{blue} layers or trilayers? \color{black}.  layers 

    %%%%% FIGURE 2 VERSION 1 

    %%%% FIGURE 2 VERSION 2 

    % \begin{figure}
    % \includegraphics[width=0.45\textwidth]{RuCl3 Figures No Line/Fig2_2.pdf}\caption{\textbf{Fabricated device and Experimental Setup}\\ 100x optical (a) and atomic force microscopy (AFM) (b) images of the completed $\alpha$-RuCl$_3$ device. Inset shows measured height of $\alpha$-RuCl$_3$ nanoflake used for the device alongside a nearby, thicker nanoflake for comparison. (c) Circuit schematic of the guarded force-voltage-measure-current setup used in this experiment making use of triax lines, wired to a custom probe with a copper sample enclosure where $R_1$ is the leakage resistance between the center line and inter shield, $R_2$ is the leakage resistance between the inner shield and outer shield, and $I_\text{M}$ is the current measured across the sample. (d). (e) Schematic showing the connection of the guard to the sample enclosure.} 
    % %This further isolates the measured current from leakage currents through the enclosure to ground as well as from electromagnetic interference. 
    % \label{fig:figure2}
    % \end{figure}
    The finished device (Fig. \ref{fig:figure2}(a)), was then mounted to a custom triax probe (Fig. 2(d)) to be later measured in a closed-cycle cryostat. The use of triax lines allows for guarded measurements to improve the signal to noise ratio (SNR) as experiments are performed at low temperatures where $\alpha$-RuCl$_3$ crystals are highly insulating. As the sample resistance at low temperatures is comparable to the insulation of the lines, leakage currents have the potential to corrupt the signal from the sample. The triax lines are similar to traditional coaxial lines with the addition of a second shield between the inner core and outside shield. In a guarded configuration, this inside shield is connected to one of the ammeter terminals at the level of the instrument (see Fig. \ref{fig:figure2}(c)) playing the role of a guard and isolating the signal from the sample. The voltage drop across the ammeter is very small so the inside shield is held at approximately the same potential as the input HI terminal (inner core), resulting in negligible leakage currents between the inner core and inside shield of the triax cable. Any leakage from the inside shield to the outside shield occurs in a loop isolated from the ammeter, so the signal measured from the sample is unaffected. If the inside shield were not present, leakage currents would come directly from the input HI going through R$_2$, corrupting the signal measured by the ammeter. %\color{blue} check schematics \color{black} 
    Additionally, the guard terminal of the instrument is connected to a copper sample enclosure (seen next to the probe in Fig. 2(d)), which serves to further isolate the sample from electromagnetic interference, where any current noise generated by surrounding AC or DC fields or coupling capacitances is kept away from the sample (see Fig. \ref{fig:figure2} e). 
    
    Current-Voltage (I-V) characteristics of the $\alpha$-RuCl$_3$ device were measured using an electrometer (Keithley 6517B), 
    %connected in a guarded configuration to make use of the triax lines of the probe. 
    across a temperature range from 1.45\,K to 130\,K, shown in Fig. 3(a) and 3(b), and in the presence of out-of-plane magnetic fields up to 11T. Current values at fixed biases were taken from each curve and fit to different transport mechanisms dependent on the temperature and bias voltage regime as outlined in Fig. 3(c). The electrodes used in this measurement are the top two depicted in Fig. \ref{fig:figure2} (b) with edge width \SI{1.0}{\micro\meter} and spacing from \SI{1}{\micro\meter} at the widest to \SI{200}{\nm} at the narrowest.
    
\section{Results}
    \begin{figure*}
    \includegraphics[width=\textwidth]{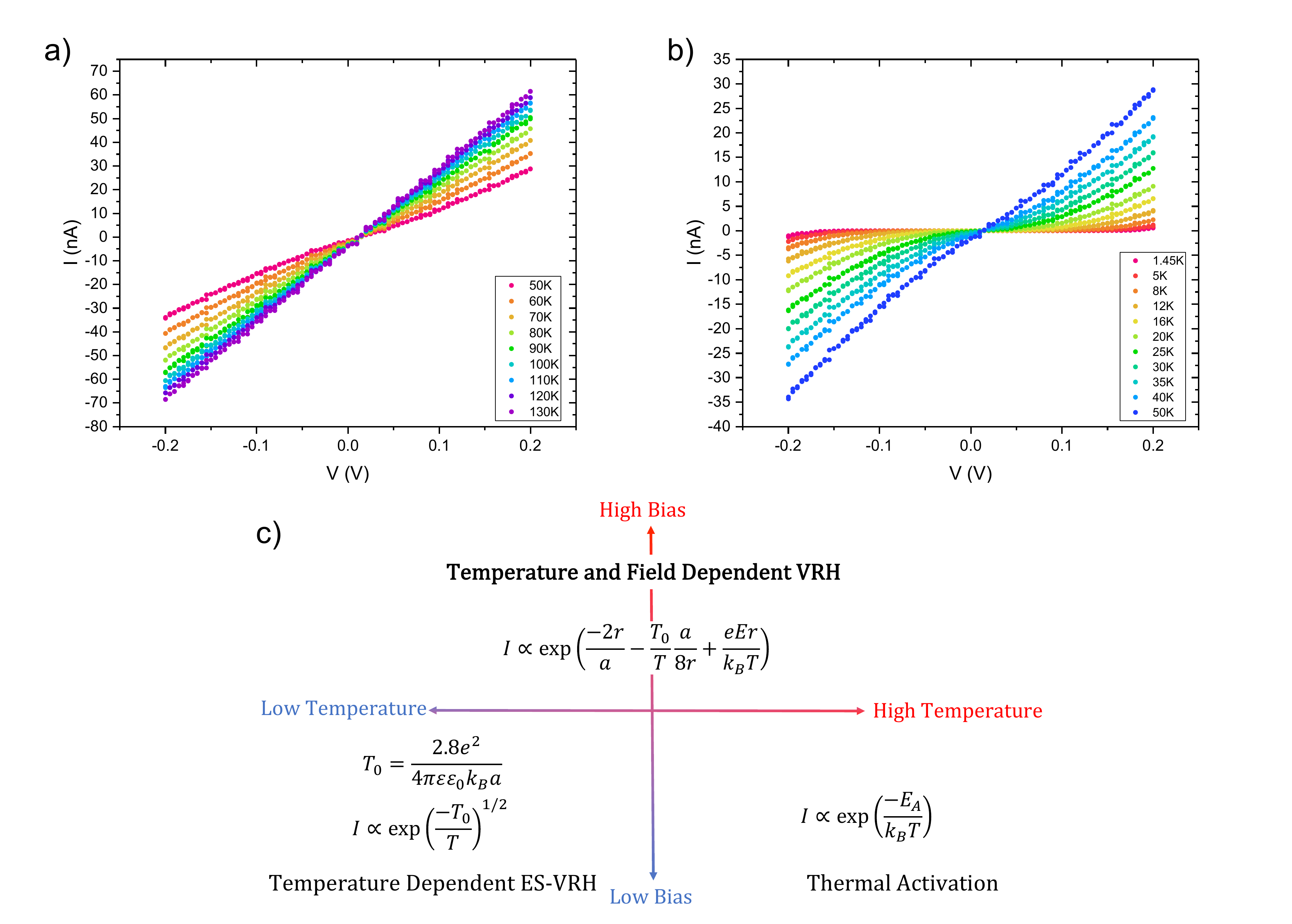}\caption{\textbf{I-V Dependence and Transport Mechanisms in Different Regimes}\\
    Current-Voltage characteristics for $\alpha$-RuCl$_3$ device showing a (a) linear, ohmic behavior at high temperatures and a (b) nonlinear behavior at low temperatures. The decreasing slope of the curves with decreasing temperatures is characteristic of a highly insulating behavior. (c) Schematics of the dominant transport mechanisms at the different regimes. Low temperatures ($<30$\,K) and low bias (\SI{25}{\mV}) are ruled by Efros-Shklovskii variable range hopping (bottom-left quadrant), high temperatures and low bias by a thermal activation mechanism (bottom-right quadrant), while high bias is dominated by an electric field-assisted hopping (top- right and -left quadrants).
    }
    \label{fig:figure3}
    \end{figure*}    
    Overall, the measured I-V curves showed an ohmic behaviour at \SI{130}{\kelvin} with increasing non-linearity and decreasing slope as the temperature is lowered, reaffirming the Mott insulating character of the material\cite{zhou2016angle, PhysRevB.90.041112} as represented in Fig. \ref{fig:figure3} (a) and (b). We then represented the data in an appropriate semi-log scale in accordance with the regimes presented in Fig. \ref{fig:figure3}(c). In particular, we found that for low bias voltage and low temperatures, the data best supports temperature dependent Efros-Shklovskii variable range hopping. At low bias and high temperatures, thermal activation dominates while for high bias, the sample is an intermediate regime as shown in Fig. \ref{fig:figure3}.

\begin{figure*}
    \includegraphics[width=\textwidth]{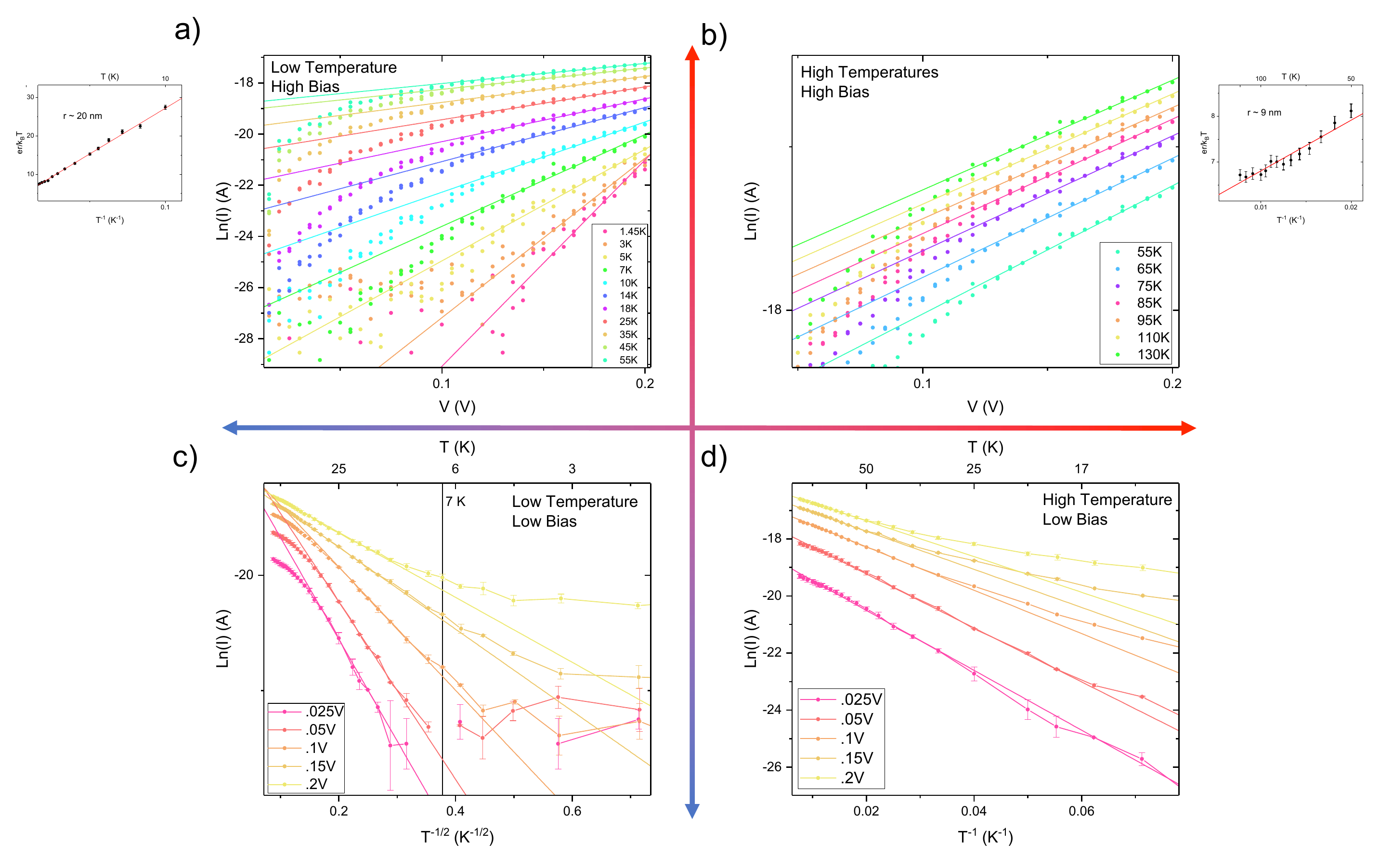}
    \caption{\textbf{Electronic Transport Mechanisms in Four Different Regimes}\\
    % \color{blue} Add a), b), c), d) labels to the figure\color{black} 
    A schematic that depicts four different regimes segmented by temperature and bias where data is represented as a semi-log plot of $\ln(I)$ in accordance with the transport mechanism dominant in that regime, according to Fig. \ref{fig:figure3}(c). In (a) and (b), for high $V$ and low and high $T$ respectively, conduction is dominated by field assisted ES-VRH. Insets show the mean hopping length, r, deduced from the linear fits shown in (a) and (b). Instead for low $T$ and low $V$, temperature dependent ES-VRH instead fits best (c). For high $T$ and low $V$, a thermally activated nearest-neighbor hopping mechanism is more prevalent. 
    } 
    \label{fig:figure4}
    \end{figure*}

    At higher temperatures (30 - 130\,K) and low bias (25\,mV) we observe that transport is thermally activated and fits to an Arrhenius law. That is, the current $I$ has a dependence on the temperature of the form 
    \begin{equation}\label{ThermalAct}
        I \propto \exp\left(\frac{-E_A}{k_B T}\right)
    \end{equation}
    where $E_A$ is the activation energy and $k_B$ is Boltzmann's constant. 
    %In this regime, thermal energy excites charge carriers to hop to nearest-neighbor sites. 
    %Then by fitting to the linearized data as in a semi-log plot as $\ln(I)$ versus $T^{-1}$ such as in 
    By fitting the data to eqn. \ref{ThermalAct} (Fig. \ref{fig:figure4} (d)), we extract an activation energy $E_A  \approx \SI{9}{\meV} $ at \SI{25}{\mV} bias voltage, that decreases with increasing bias voltage, because the carrier activation is facilitated by the increasing in-plane electric field.  As a consequence, as observed in Fig.\ref{fig:figure4}(d), the temperature range in which the data fits thermal activation shrinks at higher bias voltages. 
    %As the temperature decreases in the presence of an in-plane electric field, it is no longer energetically favorable for carriers to hop to nearest-neighbors. 

    The activation energy that we deduce $E_A$,  is several orders of magnitude smaller than the energy band gap that we observed through ARPES  for bulk crystals of $\alpha$-RuCl$_3$ (at least 1.2\,eV) and reported in the literature \cite{zhou2016angle,Sinn2016}. We attribute this to impurity states induced in the gap via charge trapping in the SiO$_2$ substrate, forming carrier puddles\cite{ESBookch10.1, JETPHuangY} as has been found for atomically thin semiconductors such as MoS$_2$ \cite{doi:10.1021/nn202852j}. Due to the size of the gap, fluctuations in the trapped charges at the interface between the crystal and the substrate should not be effectively screened at lower temperatures\cite{ESBookch10.1, JETPHuangY}. 
    %\color{blue} also talk about thermal activation (percolation level, in addition to nearest neighbor hopping ) \color{black} 
    In this temperature range (30 - 130\,K), where the temperature is much smaller than the known energy gap for $\alpha$-RuCl$_3$, we believe that carriers hop between nearest neighbor electron and hole puddles, with the activation energy reflecting the puddle charging energy \cite{JETPHuangY}.
    
    Instead at low temperatures and low bias, conduction tends to be more appropriately described by a variable range hopping (VRH) law
    % \begin{equation}
    %     I \propto \exp\left(\frac{T_0}{T}\right)^{\nu}
    % \end{equation} 
    % should there be a minus sign here?
        \begin{equation}\label{VRH}
        I \propto \exp\left(-\left(\frac{T_0}{T}\right)^{\nu}\right)
    \end{equation} 
    where $T_0$ is the characteristic temperature whose energy scale is related to the localization length of the charge carriers. Taking $\nu = 1/3$ and $\nu = 1/4$ gives Mott's VRH law for 2D and 3D hopping conduction, respectively. Mott's VRH law describes hopping of carriers between remote sites in a narrow band near the Fermi level\cite{doi:10.1080/14786436908216338}. Bulk $\alpha$-RuCl$_3$, has a 1.2 - \SI{1.9}{\electronvolt} Mott insulating gap with a zero density of states near the Fermi level as reported in the literature\cite{zhou2016angle,Sinn2016} and corroborated with our ARPES measurements (Fig. \ref{fig:figure1}(c))) which disfavors Mott VRH as a possible transport mechanism for $\alpha$-RuCl$_3$. It tuns out however that in our thin crystal device, the presence of charged impurities creates a narrow band of localized states near the Fermi level, and we believe that further electronic correlations between these impurities open a Coulomb gap. In this case, transport is described by Efros and Shklovskii's $\nu = 1/2$ variable range hopping (ES-VRH)\cite{ESBookch10.1, ALEfros_1975} 
    %that holds for both 3D and 2D hopping conduction
    % \begin{equation}
    %     I \propto \exp\left(\frac{T_0}{T}\right)^{1/2}
    %     ,
    %     \;T_0 = \frac{2.8 e^2}{4 \pi \varepsilon \varepsilon_0 k_B a}
    % \end{equation}
    \begin{equation}
        I \propto \exp\left(-\left(\frac{T_0}{T}\right)^{1/2}\right)
        ,
        \qquad T_0 = \frac{2.8 e^2}{4 \pi \varepsilon \varepsilon_0 k_B a} 
        \label{eqn:T0}
    \end{equation}
    where $e$ is the electron charge, $\varepsilon$ is the dielectric constant of the material, $\varepsilon_0$ is the permittivity of free space, and $a$ is the localization length of the impurities. For $\alpha$-RuCl$_3$ the dielectric constant is $\sim$ 7.4 \cite{Reschke_2018}. By plotting the log of the current as a function of $T^{-1/2}$, we find that ES-VRH fits well our data for low temperatures and low biases. 
    %in the temperature range 10-30\,K at a 
    For the lowest bias voltage used (25\,mV) we observe that the ES-VRH mechanism rules the electronic transport in our sample down to the zig-zag antiferromagnetic ordering transition at $\sim$7\,K as shown in Fig. \ref{fig:figure4} (c). 
    Moreover, by comparing the fitting to 3D Mott variable range hopping, we find that ES-VRH provides a slight better fit to our data, as shown in Fig. \ref{fig:supplemental} in the supplemental materials. 
    %As observed in Fig. \ref{fig:supplemental} (b), when fitted to a 3D Mott variable range hopping model, the data does not exhibit a large enough range of temperatures that readily supports linear fits. 
    %From this observation and the known vanishing density of states at the Fermi level that derive from electronic correlations and spin-orbit coupling in $\alpha$-RuCl$_3$, we conclude that transport at low temperatures and low biases in this material is most likely ruled by ES-VRH. 
    From our semi-log plot, a linear fit of the data allows us to extract $T_0$ deducing a localization length $a \approx$ \SI{3}{\nano\meter} (eqn. \ref{eqn:T0}). 
%    Then, from the definition of $T_0$ (eqn. \ref{eqn:T0}) , we are able to deduce a localization length, $a \approx$ \SI{3.2}{\nano\meter}. 
This is a similar value to the one we deduced in the past for another spin-orbit assisted Mott insulator (Na$_2$IrO$_3$)\cite{PhysRevB.101.235415} where variable range hopping is mediated by a quasiparticle at the Fermi level that results from the Ir-O octahedra embedded between two Na layers creating a charge transfer from Na to Ir in Na$_2$IrO$_3$ \cite{PhysRevB.101.235415, PhysRevB.96.161116}. 
%where Ir-O octahedra embedded between two Na layers create a charge transfer from Na to Ir leading to a quasiparticle at the Fermi level, observable through ARPES, that mediates Mott's three dimentional VRH
%and what has been reported for bulk $\alpha$-RuCl$_3$
Additionally, VRH has been reported in thin crystals of $\alpha$-RuCl$_3$ \cite{doi:10.1021/acs.nanolett.8b00926}. Although the origin of a non-zero density of states at the Fermi level is not commented in ref. \onlinecite{doi:10.1021/acs.nanolett.8b00926}, it may have the same origin as in our samples; charged impurities from the SiO$_2$ substrate. Transport data in that work may also be fitted to ES-VRH yielding similar characteristic energies to those in our samples.
    
    At high bias voltages (up to \SI{0.2}{\volt}), the energy difference in hopping sites is compensated by the electric field. Following Shklovskii's theory of hopping conduction in the presence of a strong electric field \cite{shklovskii_1973}, we use a engineered expression for intermediate fields introduced in ref. \onlinecite{PhysRevLett.92.216802},  
    \begin{equation}\label{Intermediate}
        I \propto \exp\left(-\frac{2r}{a} - \frac{T_0}{T}\frac{a}{8r} + \frac{erE}{k_BT}\right).
    \end{equation}
    % should this be the same equation as in Fig. 3(c) underneath the temperature and field dependent VRH? 
    % \begin{equation}
    %     I \propto \exp\left(\frac{2r}{a}+\frac{T_0}{T} - \frac{erE}{k_BT}\right)
    % \end{equation}
    Here, the Miller and Abrahams relation for hopping conduction \cite{ESBookch10.1} is modified to include the contribution from the field, leading for E=0 to the ES-VRH expression (eqn. \ref{VRH}). On the other hand, when the electric field exceeds a critical value, $E_C \approx  \frac{2k_B T}{e a}$, the energy required for hopping is fully compensated by the field. As a result, the second and third terms in expression \ref{Intermediate} cancel each other (see Supp Mat.), leading to a temperature independent transport mechanism \cite{PhysRevLett.92.216802}, consistent with Shklovskii's theory \cite{shklovskii_1973},
    \begin{align}\label{E}
        I \propto \exp \left(-\left(\frac{E_0}{E}\right)^{1/2}\right),
        \qquad E_0 = \frac{k_B T_0}{2ea}. 
    \end{align}
    Indeed, eqn. \ref{Intermediate} is in reasonable agreement with numerical results reported for fields below $E_C$ \cite{00d4213e6c154262a12bf80e0139add2}.  
    Using the previously found value for the localization length $a \approx $ \SI{3}{\nano\meter}, we estimate the critical field to be $E_C >$ \SI{0.2}{\volt/\micro\meter} for temperatures above the 7\,K magnetic ordering transition. 
    %Calculating this critical field $E_C$ with localization length $a \approx $ \SI{3.2}{\nano\meter} deduced from the low bias ES-VRH regime, we see that $E_C >$ \SI{0.2}{\volt/\micro\meter} for temperatures above the magnetic transition at 7\,K. 
    As our range of fields here are 0.0-\SI{0.2}{\volt/\micro\meter}, transport in this regime is not fully driven by the electric field and is dependent on both the temperature and the field, following therefore eqn. \ref{Intermediate}. A semi-log plot of $I$ vs $V$ allows us to extract the temperature dependence of the third term in eqn. \ref{Intermediate} by fitting each of our I-V curves (see Fig. \ref{fig:figure4}a). By analyzing this temperature dependence, we acquire the mean hopping length, $r$ (see insets Fig. \ref{fig:figure4} a and b). In a temperature range from 130\,K to 55\,K we calculate $r\approx$ \SI{9}{\nano\meter}, while in a lower range of temperatures of 55\,K to 10\,K, $r$ is instead $\approx$ \SI{20}{\nano\meter}. We observe therefore a shorter hopping distance at higher temperatures, indicating that as temperature increases, carriers are more likely to hop to spatially neighboring sites. A longer hopping length at lower temperatures is consistent with variable range hopping mechanism as carriers look to more distant sites to find energetically favorable hops. Such behavior can be seen directly in the data, in the temperature dependence of the slopes of the fitted data in Fig. \ref{fig:figure4}a (field-assisted regime). Above a temperature of $\approx$ 55\,K (Fig. \ref{fig:figure4}b), the slopes change very little with the temperature, indicating that the activation energy is compensated less by the field and that shorter hops are favored. As temperature decreases, the slopes increase and hops cover in average more distance, in the direction of the electric field.

    \begin{figure}
    \includegraphics[width=0.48\textwidth]{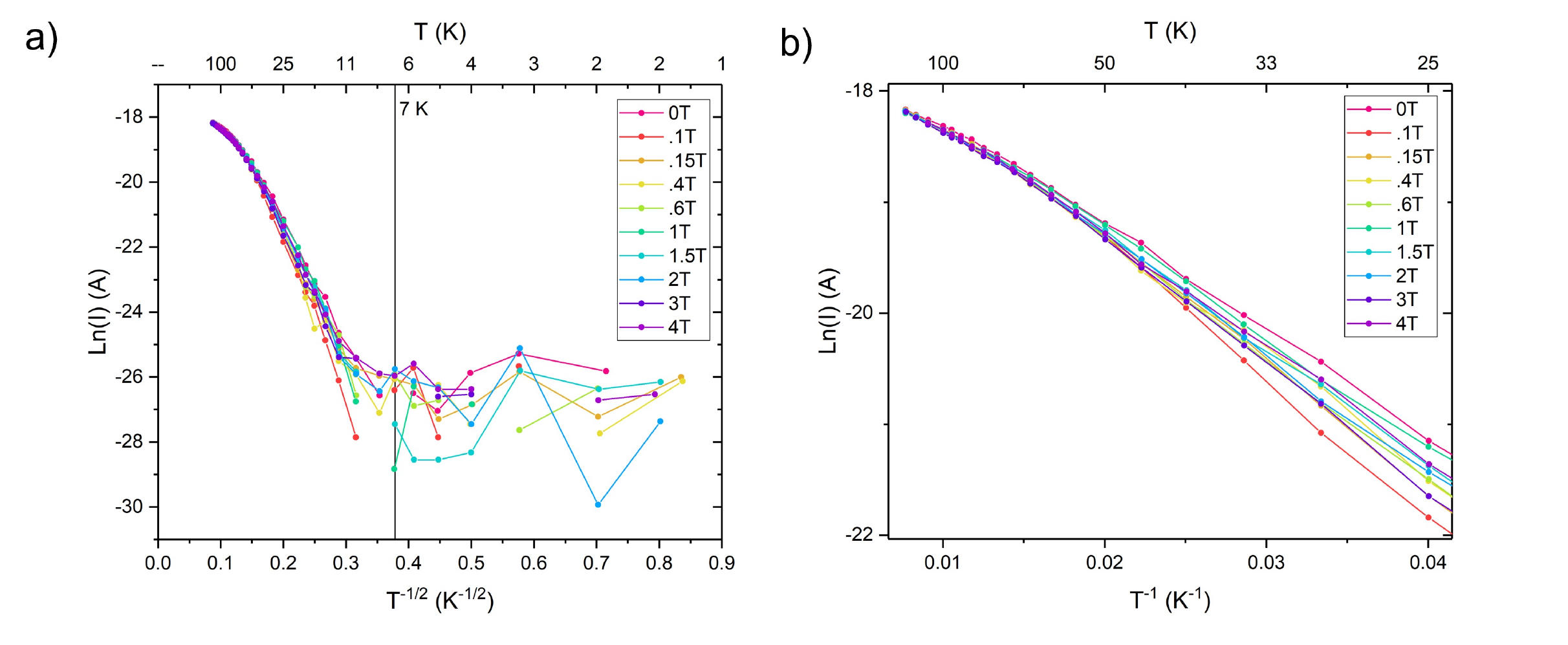}
    \caption{
    \textbf{Effect of an Out-of-Plane Magnetic Field on the transport mechanisms}\\
    Data taken from I-V curves at \SI{50}{mV} in a range of magnetic fields perpendicular to the honeycomb lattice, from \SI{0}{\tesla} to \SI{4}{\tesla} plotted as $\ln(I)$ versus $T^{-1/2}$  in fitting to Efros-Shklovskii variable range hopping (a) and as $\ln(I)$ versus $T^{-1}$ in fitting to thermal activation (b). In either case, we do not see a consistent effect of an out-of-plane magnetic field on the transport, either above or below the magnetic ordering transition (7\,K)% in line with $\alpha$-RuCl$_3$'s anisotropic magnetic field response\cite{PhysRevB.91.180401}
    }
    \label{fig:figure5}
    
    % \caption{\label{Figure3}
    % \textbf{Circular dichroism of VI$_3$ ARPES} a) VI$_3$ ARPES using left-hand circularly (LC) (a) and right-hand circularly polarized (RC) light (b) along the K-$\Gamma$-K direction. (c) Circular dichroism calculated using (a) and (b). (e) Schematics of the experimental setup as in figure \ref{Figure2}e showing incident photons right-hand (red) and left-hand (blue) circularly polarized. }
    \end{figure}

\begin{figure}
    \includegraphics[width=0.49\textwidth]{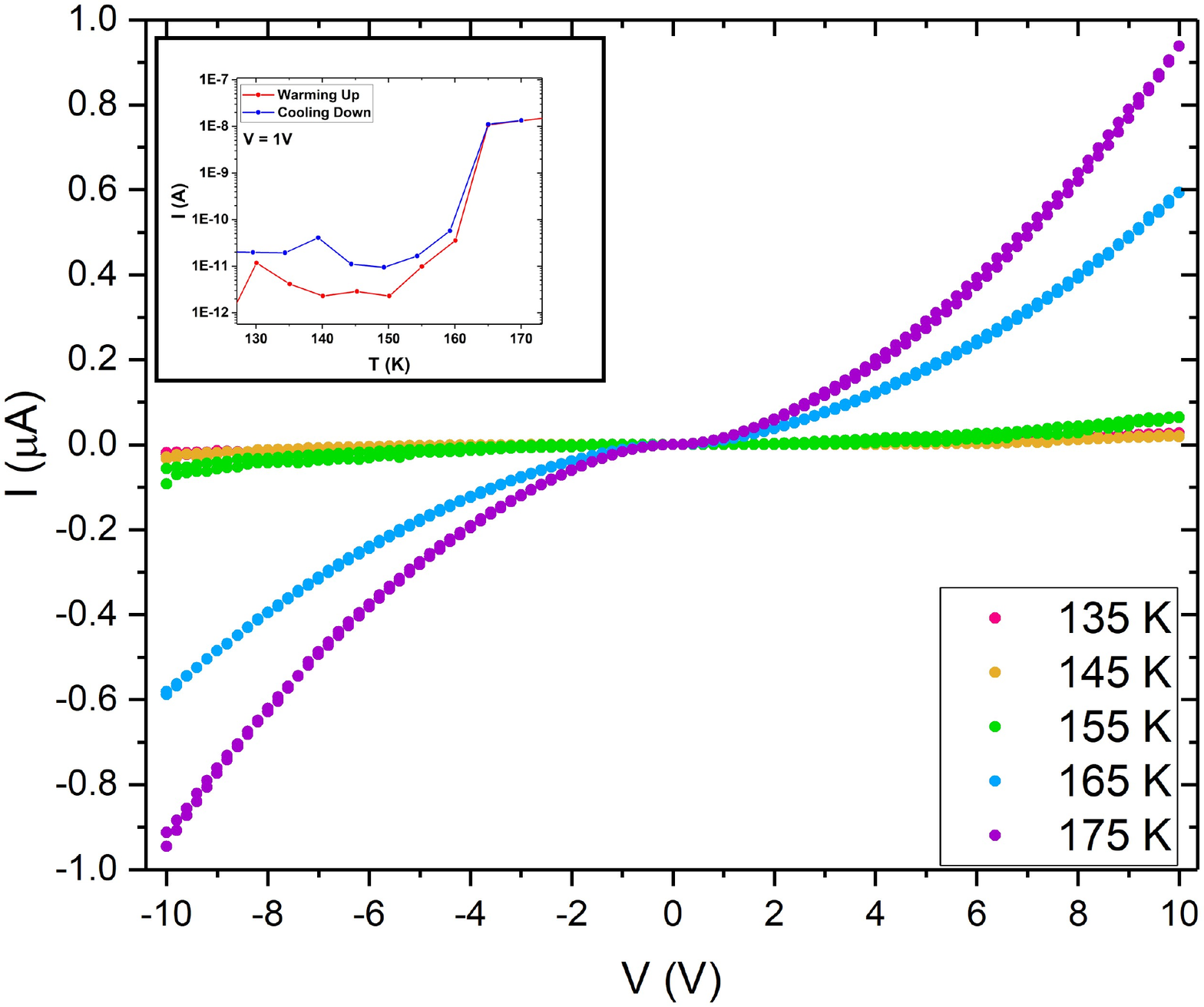}
    \caption{\textbf{Signature of Structural Phase Transition in a thin crystal of $\alpha$-RuCl$_3$}\\
    I-V characteristics of a separate thin crystal device of $\alpha$-RuCl$_3$, measured up to 175\,K, where a striking change in the conductance of the sample at \SI{160}{\kelvin} can be observed, best seen in the inset, where the current from I-V curves at different temperatures was extracted at bias voltage 1\,V. 
    %In temperature range, we were able to see signatures of the structural phase transition at $\sim$ 155\,K\cite{PhysRevB.93.134423,Ziatdinov2016} in the form of an abrupt change in the slope of the curves. Taking the current at a fixed bias of 1\,V vs T (inset) we see a hysteretic behavior comparing data taken warming up (red) to data taken cooling down (blue).
    }
    \label{fig:phaseTransition}
\end{figure}

    \par
    Our data shows that while electronic transport in a thin crystal of $\alpha$-RuCl$_3$ can be well understood in a wide range of temperatures, it goes beyond ES-VRH or thermal activation below the magnetic ordering transition at 7\,K. This result is in line with our previous work on the spin-orbit assisted insulator Na$_2$IrO$_3$ \cite{PhysRevB.101.235415}. 
    %However here, the more important insulating character of nanoflakes of $\alpha$-RuCl$_3$ demanded the use of more specialized techniques for high impedance measurements. 
    
    We noted no change in the transport mechanism near 14\,K, where a magnetic ordering transition has been reported for $\alpha$-RuCl$_3$ crystals with a two-layer AB stacking order (instead of a ABC configuration). This points to a $\alpha$-RuCl$_3$ thin crystal device free of deformations or stacking faults  \cite {PhysRevB.93.134423}. 
    %We noticed that $\alpha$-RuCl$_3$ having a zero density of states at the Fermi level, is ruled at low temperatures and low bias by ES-VRH, in contrast to Na$_2$IrO$_3$, where Ir-O octahedra embedded between two Na layers create a charge transfer from Na to Ir leading to a quasiparticle at the Fermi level, observable through ARPES, that mediates Mott's three dimentional VRH. 
    %the antiferromagnetic ordering hampers our ability to accurately measure electronic transport.
    
    We observed no change in any of the transport mechanisms in the thin flake of $\alpha$-RuCl$_3$ or the deduced transport quantities, when an out-of-plane magnetic field up to 11\,T was applied, as seen in Fig. \ref{fig:figure5}.   
    %As demonstrated in Fig. \ref{fig:figure5} (a) and (b), there was no clear magnetic field dependence shown in the transport with magnetic fields up to 11\,T. 
    This is consistent with the known magnetic anisotropy in $\alpha$-RuCl$_3$, that as mentioned in the introduction, results in a critical field that increases dramatically from the in-plane ($\approx$ 7\,T) to the out-of-plane direction ($\approx$ 33\,T)\cite{https://doi.org/10.48550/arxiv.2201.04597,PhysRevB.91.180401}.

Finally, we want to point out that in a second $\alpha$-RuCl$_3$ thin crystal sample where I-V curves were measured up to \SI{175}{\kelvin}, we observed a drastic change in the conductance at $\approx 160$\,K by almost three orders of magnitude, apparent in both cooling down and warming up cycles (see Fig. \ref{fig:phaseTransition}). We believe that this is signature of the structural phase transition reported through neutron diffraction experiments for exfoliated thin crystals of $\alpha$-RuCl$_3$ \cite{Ziatdinov2016}, and puts in evidence the reliability of our high impedance guarded measurements.
    
     % \begin{equation}
     %     I \propto \exp\left(\frac{2r}{a}\right)
     % \end{equation}

     % Using the identities $\frac{e^2}{4 \pi \varepsilon \varepsilon_0 r} = e E r$ and $T_0$ from (3) taken at a low field, we arrive at a field-dependent expression for the current due to hopping conduction

     % \begin{equation}
     %     I \propto \exp\left(\frac{E_0}{E}\right)^{1/2}
     % \end{equation}

     % where $E_0 = \frac{k_B T_0}{2 e a}$ is the characteristic electric field. 

 \section{Conclusion}
    In summary, we fabricated an electronic device of a mechanically exfoliated $\alpha$-RuCl$_3$ nanoflake. The use of guarding and triax lines allowed us to isolate the high impedance input of our measuring instrument (an electrometer) from leakage currents in the experimental setup, granting us access to electronic transport measurements in our $\alpha$-RuCl$_3$ devices at temperatures as low as 1.5\, K. Additionally, the use of a custom-made metallic shielding around our sample allowed us to reduce the noise associated to electrostatic interference. With the help of these techniques, we found that at low bias voltages, transport is thermally activated at high temperatures (30-\SI{130}{\kelvin}) 
    % \textcolor{red}{what temperature range?} 
    and follows Efros-Shklovskii (ES) variable range hopping at low temperatures 
    % \textcolor{red}{what temperature range?}
    ($\approx$7-\SI{30}{\kelvin}). With increasing bias, the transport mechanism changes to a field-assisted  one, but is not entirely field driven within the voltage range used in this experiment. From these different models, we deduced the presence of charged impurities, possible originated at the SiO$_2$/RuCl$_3$-thin crystal interface, with a localization length of $a \approx \SI{3}{\nm}$. We observed mean hopping lengths at different temperature ranges in the field-assisted regime, following the expected behavior for variable range hopping at lower temperatures, where charge carriers hop to remote sites with close energy levels. Most importantly, we found that ES-VRH is suppressed below the zigzag antiferromagnetic ordering transition temperature known for bulk $\alpha$-RuCl$_3$, 7\,K. This behavior is unaffected by the presence of magnetic fields in the c-axis (perpendicular to the honeycomb plane) as large as 11\,T, consistent with recent results that suggest the suppression of the magnetic ordered state in $\alpha$-RuCl$_3$ at out-of-plane fields larger than 33\,T, leading to a phase transition to a QSL \cite{https://doi.org/10.48550/arxiv.2201.04597}. Finally, we observed signature of the structural phase transition reported for thin crystals of $\alpha$-RuCl$_3$\cite{Ziatdinov2016}, which testifies the fidelity of our technique to measuring high impedance samples.  Future experiments will take advantage of the strong magnetic anisotropy in $\alpha$-RuCl$_3$, using an in-plane magnetic field, for which the critical field is considerably reduced \cite{Kasahara2018, Czajka2021}, enabling the study of predicted signatures of a QSL state through electronic transport experiments at low temperatures \cite{PhysRevResearch.2.033439}.  

    %To do 
    %Check captions
    %Add acknoledegments Patrick, Vinh
    
    %Measurements  an out-of-plane magnetic field. The transport mechanism below the zig-zag antiferromagnetic transition goes beyond both ES-VRH and the field-assisted model. Future experiments will apply an in-plane magnetic field in order to suppress this ordering\cite{Kasahara2018, Czajka2021} and explore the electronic transport at low temperatures. This work demonstrates the efficacy of guarding in high impedance measurements to explore electronic transport in insulating materials.

\section*{Supplementary Material}
Supplemental materials includes a figure comparing the fits of Efros-Shklovskii variable range hopping versus 3D Mott variable range hopping in the low temperature and low bias regime (Fig. \ref{fig:supplemental}). Also included are calculations demonstrating that the expression describing transport mechanisms in intermediate regimes, eqn. \ref{Intermediate}, reduces to temperature dependent ES-VRH hopping $\sim \exp(-(T_0/T)^{1/2})$ (eqn. \ref{eqn:T0}) in the limit of low temperatures and low fields and to field assisted ES-VRH hopping in the limit of strong fields $\sim \exp(-(E_0/E)^{1/2})$ (eqn. \ref{E}).

% It also shows the mathematical deduction for expression \ref{E}.

% \subsection*{Field driven transport mechanism}
% The electric field driven transport mechanism presented in eqn. \ref{E} of the main text, derives from eqn. \ref{Intermediate}, when the field reaches a critical value $E_C \approx  \frac{2k_B T}{e a}$. For $E>E_C$, 
% \begin{align}\nonumber
%     G = A \exp\left(-\sqrt{\frac{E_0}{E}}\right) \qquad E_0 = \frac{k_BT}{2ea}
% \end{align}
% which is equivalent to 
% \begin{align}\nonumber
%     G  = B\exp\left(-2\frac{r}{a}\right)
% \end{align}
% where A and B are real constants. This is deduced by writing $r$ in terms of the electric field $E$,
% \begin{align}\nonumber
%     E = \frac{e}{4\pi\epsilon_0 \epsilon r^2}
% \end{align}
% and the localization length $a$ in terms of $T_0$
% \begin{align}\nonumber
%     T_0 \approx \frac{2.8 e^2}{4\pi\epsilon_0 \epsilon a k_B}.
% \end{align}

\subsection*{Low Temperature, Low Field Reduction to ES-VRH}
The intermediate electric field and temperature driven transport mechanism as described in eqn. \ref{Intermediate} in the main text can be shown to reduce to ES-VRH where the conductivity goes as $\exp(-(T_0/T)^{1/2})$ in the limit of low temperatures and low fields \cite{ALEfros_1975}. In this case, the third term dependent on $E$ vanishes and we are left with
\begin{align}
    G \propto \exp\left(-\frac{2r}{a} - \frac{T_0}{T}\frac{a}{8r}\right) \label{eqn:G_low_temp}
\end{align}
Then we would like to find an optimum $r$ such that $G$ is maximized. We can differentiate this expression with respect to $r$ and set it equal to zero
\begin{align}
    \frac{dG}{dr} &= 
    A\left[-\frac{2}{a} + \frac{T_0}{T}\frac{a}{8 r^2}\right]\exp\left(-\frac{2r}{a} - \frac{T_0}{T}\frac{a}{8r}\right) = 0
\end{align}
Solving for $r$ gives us
\begin{align}
    r = \frac{a}{4}\sqrt{\frac{T_0}{T}}
\end{align}
Inserting this expression for $r$ back into eqn. \ref{eqn:G_low_temp} then reduces it to the $\exp(-\sqrt{T_0/T})$ dependence that is characteristic of ES-VRH \cite{ESBookch10.1, ALEfros_1975}.
% \begin{align}
%     G &= A\exp\left(-\frac{2}{a}\cdot \left(\frac{a}{4}\sqrt{\frac{T^*}{T}}\right) - \frac{T^*}{T}\frac{a}{8}\cdot \left(\frac{4}{a}\sqrt{\frac{T}{T^*}}\right)\right)\\
%     &= A\exp\left(-\frac{1}{2}\sqrt{\frac{T^*}{T}} - \frac{1}{2} \sqrt{\frac{T^*}{T}}\right)\\
%     &= A\exp\left(-\sqrt{\frac{T^*}{T}}\right)
% \end{align}
\begin{align}
    G &= A\exp\left(-\sqrt{\frac{T_0}{T}}\right).
\end{align}

\subsection*{High Field Limit Reduction to Field Assisted ES-VRH}
Similarly, the intermediate electric field and temperature driven expression eqn. \ref{Intermediate} can be shown to reduce to the strong field limit expression for field assisted ES-VRH \cite{shklovskii_1973, dvurechenskii_1988}. For large $E$, the last two terms vanish in eqn. \ref{Intermediate} cancel due to the transport mechanism transitioning to an activationless regime. As a result, we can set these two terms equal to zero 
\begin{align}
    \frac{T_0}{T}\frac{a}{8r} = \frac{eEr}{k_BT}
\end{align}
and solve for $r$, yielding 
\begin{align} 
    r = \sqrt{\frac{k_B T_0 a}{8 eE}} \label{eqn:r_1}
\end{align}
Inserting eqn. \ref{eqn:r_1} back into eqn. \ref{Intermediate} then cancels the last two terms and modifies the first term to 
\begin{align}
    G = A \exp\left(- \frac{2}{a}  \sqrt{\frac{k_B T_0 a}{8 eE}}\right)  \label{eqn:G_E_mod}
\end{align}
But from the definition of $E_0$ in eqn. \ref{E}, we obtain an expression for $a = a(E_0)$  
\begin{align}
    a = \frac{k_BT_0}{2eE_0}
\end{align}
that can inserted into eqn. \ref{eqn:G_E_mod} to show that $G$ reduces to the expression for ES-VRH in the limit of strong $E$ fields \cite{shklovskii_1973}. 
\begin{align}
    G  = 
    A\exp\left(-\sqrt{\frac{E_0}{E}}\right)
\end{align}
% \begin{align}
%     G &= A\exp\left(-2 \sqrt{\frac{2eE_0}{k_BT_0}}\sqrt{\frac{k_BT_0}{8eE}}\right)\\
%     &= A\exp\left(-2\sqrt{\frac{2E_0}{8E}}\right)\\
%     &= 
%     A\exp\left(-\sqrt{\frac{E_0}{E}}\right)
% \end{align}
% which is the expression for ES-VRH in the limit of strong fields. 

\subsection*{Data fit to Efros-Shklovskii (ES-VRH) and three-dimensional variable range hopping (3D-VRH)}
\begin{figure}
    \includegraphics[width=0.48\textwidth]{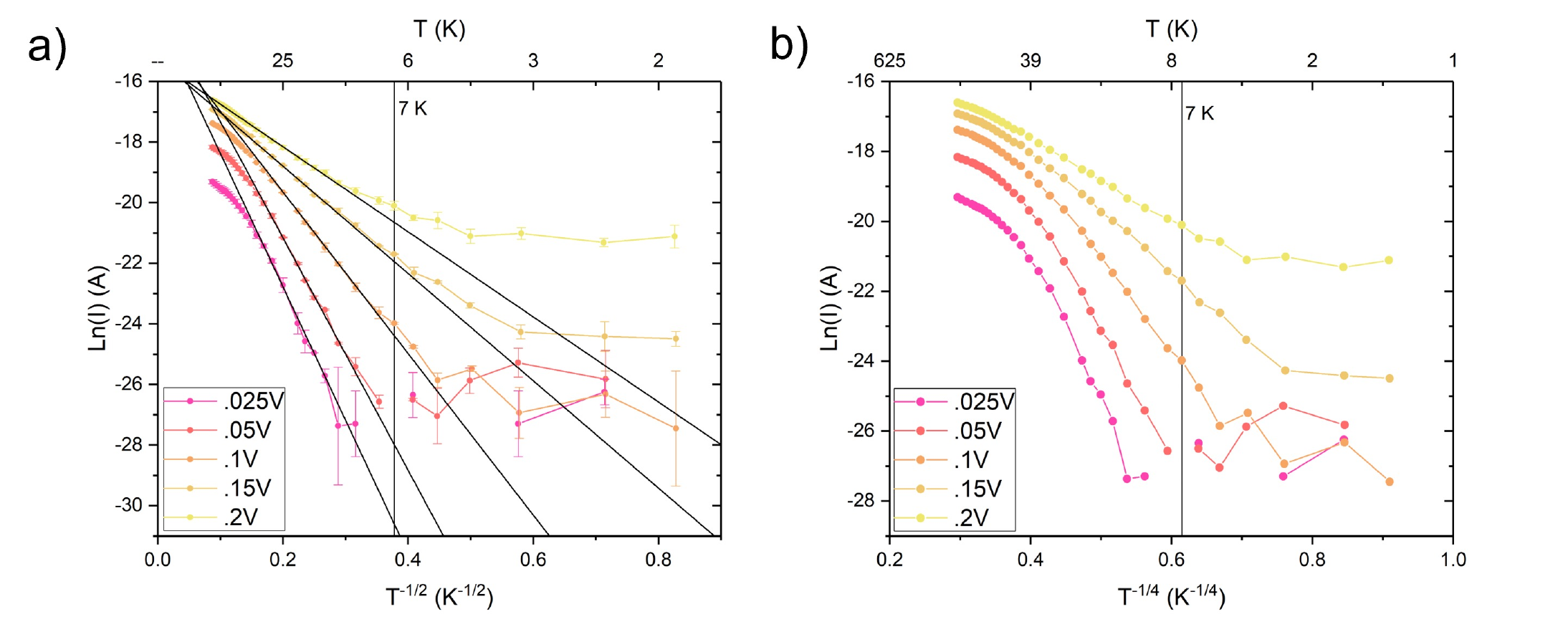} 
    \caption{\textbf{ES-VRH vs. 3D Mott VRH}\\
    Comparison of Efros-Shklovskii variable range hopping (a) versus 3D Mott variable range hopping (b). ES-VRH fits our data in a slight wider temperature range.
    }
    \label{fig:supplemental}
\end{figure}

Data comparing Efros-Shklovskii variable range hopping to 3D Mott variable range hopping is presented in Fig. \ref{fig:supplemental}. 

\begin{figure}
    \includegraphics[width=0.49\textwidth]{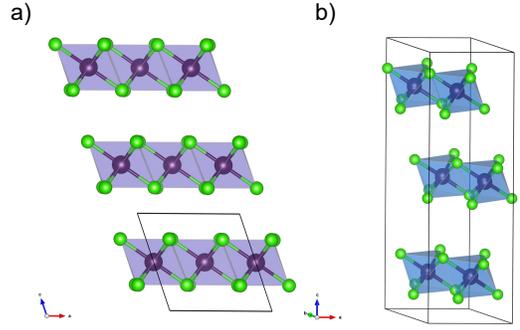}
    \caption{\textbf{C2/m (a) and P3$_1$12 (b) crystal structures of $\alpha$-RuCl$_3$.}\\
    Outlined are the unit cells of these configurations. Monoclinic C2/m structure shows a one-layer unit cell with an ABC stacking order while the P3$_1$12 rhombohedral structure contains a three-layer unit cell with an AB stacking.
    }
    \label{fig:supplemental2}
\end{figure}

\begin{acknowledgments}
The primary funding for this work was provided by the U.S. Department of Energy, Office of Science, Office of Basic Energy Sciences under contract DE-SC0018154.  The crystal growth work by V.N. and J.G.A. was supported by the National Science Foundation under Grant No. 1905397.
Atomic Force Microscopy imaging was supported by the National Science Foundation, MRI program, under award number NSF-MRI 2018653. The photoemission work by A.L. was supported by the Novel sp2-Bonded Materials and Related Nanostructures Program (KC2207). The Advanced Light Source is supported by the Director, Office of Science, Office of Basic Energy Sciences, of the U.S. Department of Energy (U.S. DOE-BES) under contract no. DE-AC02-05CH11231. We would like to acknowledge invaluable advice from Prof. Boris Shklovskii regarding our theoretical treatment of intermediate regimes, as well as Jonathan Denlinger from the Advanced Light Source at LBL, and David Warren from Oxford Instruments for advice on our triax lines probe and Nicholas Breznay for advice on high impedance measurements.
P.B. would like to acknowledge the Margaret Heeb Summer Research Assistantship in Honor of Wilma Jordan
Olaf and the Mary Jane Anfinson Endowed Scholarship at CSULB. P.B. and V.T. would like to acknowledge the Richard D. Green Graduate Research Fellowship from the College of Natural Sciences and Mathematics at CSULB. M.M would like to acknowledge the Google American Physical Society Inclusive Graduate Education Network Bridge Fellowship Program at CSULB.   
\end{acknowledgments}
\bibliography{Ojeda-AristizabalBibRuCl3}

\end{document}